%% file: conference_101719.tex
\documentclass[conference]{IEEEtran}
\IEEEoverridecommandlockouts
\usepackage{array}
\newcolumntype{P}[1]{>{\centering\arraybackslash}p{#1}}
\newcolumntype{M}[1]{>{\centering\arraybackslash}m{#1}}
\usepackage{tikz}
\usepackage{amssymb}
\usepackage{xcolor}
\usepackage{url}
\usepackage{threeparttable}
\usepackage{balance}

\usepackage{hyperref}
\usepackage[htt]{hyphenat}
\newcommand{\greencheck}{{\color{black}\checkmark}}

\usepackage{algpseudocode}

\newcommand{\sof}{\textit{Stack Overflow}\xspace}
\newcommand{\gh}{\textit{GitHub}\xspace}
\newcommand{\keras}{\textit{Keras}\xspace}
\newcommand{\etal}{{\em et al.}\xspace}

\newcommand{\tabref}[1]{Table~\ref{#1}}
\newcommand{\fignref}[1]{Figure~\ref{#1}}

\usepackage[ruled,linesnumbered]{algorithm2e}

\usepackage[figuresright]{rotating}
\usepackage{multirow, array}
\usepackage{graphics}

\usepackage{graphicx,caption,subfigure}
\usepackage{amsmath,amsfonts,amssymb}
\usepackage{epstopdf}
\usepackage{hyperref}
\usepackage{lscape}
\usepackage{bbding}
\usepackage{pifont}
\usepackage{wasysym}
\usepackage{amssymb}
\usepackage{colortbl}
\usepackage{adjustbox}

\SetKw{KwStep}{Step}
\usepackage{cite}
\usepackage{amsmath,amssymb,amsfonts}
\usepackage{graphicx}
\usepackage{textcomp}
\usepackage{xcolor}
\def\BibTeX{{\rm B\kern-.05em{\sc i\kern-.025em b}\kern-.08em
    T\kern-.1667em\lower.7ex\hbox{E}\kern-.125emX}}

\usepackage{listings}
\usepackage{parcolumns}
\usepackage{url}

\usepackage{listings}
\definecolor{codegreen}{rgb}{0,0.6,0}
\definecolor{codegray}{rgb}{0.5,0.5,0.5}
\definecolor{codepurple}{rgb}{0.58,0,0.82}
\definecolor{backcolour}{rgb}{0.95,0.95,0.92}
\definecolor{Gray}{gray}{0.1}
\lstdefinestyle{mystyle}{
	backgroundcolor=\color{backcolour},
	basicstyle=\footnotesize{0.5},
	commentstyle=\color{codegreen},
	keywordstyle=\color{magenta},
	numberstyle=\tiny\color{codegray},
	stringstyle=\color{codepurple},
	basicstyle=\scriptsize,
	breakatwhitespace=false,
	breaklines=true,
	captionpos=b,
	keepspaces=true,
	numbers=left,
	numbersep=5pt,
	showspaces=false,
	showstringspaces=false,
	showtabs=false,
	tabsize=6
}

\lstdefinelanguage{Pythonna}{%
	language     = python,
	morekeywords = {to_categorical, flow_from_directory,pad_sequences,load_image}
	basicstyle=\footnotesize,
	numbers=left,
	stepnumber=1,
	showstringspaces=false,
	tabsize=1,
	breaklines=true,
	breakatwhitespace=false,
}
\lstdefinestyle{customc}{
	belowcaptionskip=1\baselineskip,
	basicstyle=\fontsize{1}{1}\selectfont\small,
	breaklines=false,
	frame= single,
	numbers = left,
	breaklines = true,
	xleftmargin=\parindent,
	language= Pythonna,
	showstringspaces=false,
	basicstyle=\footnotesize\ttfamily,
	keywordstyle=\bfseries\color{green!40!black},
	commentstyle=\itshape\color{purple!40!black},
	identifierstyle=\color{blue},
	stringstyle=\color{codegreen},
	backgroundcolor=\color{gray!4}
}

\begin{document}

\title{DeepLocalize:~Fault~Localization~for~Deep~Neural Networks\\
}
\author{\IEEEauthorblockN{Mohammad Wardat}
\IEEEauthorblockA{\textit{Department of Computer Science} \\
\textit{Iowa State University}\\
2434 Osborn Dr\\
Ames, IA, 50011, USA \\
wardat@iastate.edu}
\and
\IEEEauthorblockN{Wei Le}
\IEEEauthorblockA{\textit{Department of Computer Science} \\
\textit{Iowa State University}\\
2434 Osborn Dr\\
Ames, IA, 50011, USA \\
weile@iastate.edu}
\and
\IEEEauthorblockN{Hridesh Rajan}
\IEEEauthorblockA{\textit{Department of Computer Science} \\
\textit{Iowa State University}\\
2434 Osborn Dr\\
Ames, IA, 50011, USA \\
hridesh@iastate.edu}
}

\maketitle
\thispagestyle{plain}
\pagestyle{plain}
  
\input{abstract}


\input{introduction}
\input{motivation}
\input{approach}

\input{evaluation}
\input{threats}

\input{related}
\input{conclusion}
\input{acknowledgment}

\balance

\bibliographystyle{IEEEtran}
\bibliography{conference_101719}

\end{document}

%% file: abstract.tex
\begin{abstract}
\label{sec:abstract}
Deep neural networks (DNNs) are becoming an integral part of most 
software systems. Previous work has shown that DNNs have bugs.
Unfortunately, existing debugging techniques don't support
localizing DNN bugs because of the lack of understanding of model 
behaviors. The entire DNN model appears as a black box. 
To address these problems, we propose an approach and a tool that 
automatically determines whether the model is buggy or not, and 
identifies the root causes for DNN errors. 
Our key insight is that historic trends in values propagated between 
layers can be analyzed to identify faults, and also localize faults.
To that end, we first enable dynamic analysis of deep learning applications:
by converting it into an imperative representation and alternatively using a
callback mechanism.   
Both mechanisms allows us to insert probes that
enable dynamic analysis over the traces produced by the DNN while it is 
being trained on the training data. 
We then conduct dynamic analysis over the traces to identify the faulty layer 
or hyperparameter that causes the error. 
We propose an algorithm for identifying root causes by  
capturing any numerical error and monitoring the model during 
training and finding the relevance of every layer/parameter on 
the DNN outcome. 
We have collected a benchmark containing 40 buggy models and 
patches that contain real errors in deep learning applications from  \sof and \gh.
Our benchmark can be used to evaluate automated debugging tools 
and repair techniques. 
We have evaluated our approach using this DNN bug-and-patch benchmark, and 
the results showed that our approach is much more effective than the existing 
debugging approach used in the state-of-the-practice \keras library.
For 34/40 cases, our approach was able to detect faults whereas the 
best debugging approach provided by \keras detected 32/40 faults.
Our approach was able to localize 21/40 bugs whereas 
\keras did not localize any faults.

\end{abstract}

%% file: introduction.tex
\section{Introduction}
\label{sec:Introduction}

Deep neural networks are a class of machine learning algorithms that have 
gained significant popularity in recent years due to their remarkable success
in tasks that defy traditional algorithm techniques. 
A deep neural network can be thought of as a graph where nodes, called {\em neurons}, 
are functions with adjustable weights.
The neurons of a DNN
are organized in layers and edges feed output from a neuron to neurons in 
the next layer, and eventually to the last layer called the output layer. 
During the training step, each training input is passed through the network
to produce output.
This output is compared to the expected output. 
The difference between the actual output and the expected output, measured 
using a function called the {\em loss function}, is then used to adjust the 
weights of the neurons in the layers using a process called {\em back propagation}. 
DNNs are utilized in various software systems to make decisions. 
Thus, software engineering for DNNs has become essential. 

To aid integration of the DNN in software systems, a number of developers have
produced industrial-strength libraries and frameworks such as 
\textit {TensorFlow} \cite{TensorFlow}, \textit{Cafe} \cite{jia2014caffe}, \textit{MXNet} \cite{chen2015mxnet}, 
\textit{PyTorch} \cite{paszke2017automatic}, \textit{Theano} \cite{al2016theano} and
\keras \cite{Keras} to assist the programmers in designing reliable 
deep learning applications.
Recent work has shown that applications that use DNN have bugs~\cite{zhang2018empirical,zhang2019empirical,islam2019comprehensive}.
Same group of researchers have also studied repair strategies for 
DNN~\cite{islam20repairing}.
Zhang \etal \cite{zhang2019empirical} describe the challenges and 
limitations in detecting and localizing the bugs in the DNN model, 
and indicate that current approaches are not effective to examine the 
state of the model at some point, like the regular programs.
Islam \etal observe that DNN bug fix patterns are distinctive 
compared to traditional bug fix patterns, and that DNN bug localization is
among the major challenges faced by developers when fixing bugs~\cite{islam20repairing}.

Despite the growing number of software debugging techniques such as 
automated bug repair \cite{le2011genprog,nguyen2013semfix}, 
fault localization \cite{jones2002visualization,baudry2006improving}, 
delta debugging \cite{zeller1999yesterday}, and 
slicing \cite{zhang2006locating}, 
these techniques are still not applicable to identify the bugs in 
the DNN models and identify the faulty statements that cause the 
problem at a particular layer in the model. 
Regular software programs and the DNN models are fundamentally 
different with respect to fault and fault detection. 
For example, regular software programs are tested by comparing 
the actual output and the expected output. 
If actual output doesn't match the expected output, 
then we consider the program has a bug. 
On the other hand, the DNN-based software has a complex structure, 
and it is learning from a training dataset. 
If the DNN produces incorrect classification during training, 
we call it failure case, it is not necessarily that DNN
contains a bug, because a DNN model cannot guarantee 100\% correct
classifications. 
Furthermore, the logic of a regular program is represented in terms 
of control flow, while DNN programs use weights
between neurons and different kinds of activation functions for
similar purposes. These differences make debugging and testing of
software that deploys DNNs challenging.

Traditional practices for debugging uses aides such 
as print statements, breakpoints, and tracing the failing test case. 
These manual debugging processes take a long time and effort from 
developers~\cite{vessey1985expertise}. 
Researchers have proposed several 
automated fault localization techniques~\cite{renieres2003fault, zeller2002isolating, jones2002visualization}. 
These techniques are used to locate the root causes and understand 
the faulty states. 
Unfortunately, current automated fault localization techniques cannot 
be applied directly to DNN since existing techniques are not able to 
identify plausible and distinct root causes for unexpected behavior 
(known as failure) in DNNs. 

To overcome these challenges, this work introduces a white box 
based fault localization technique for DNNs. 
Our approach requires the source code of the DNN model and the 
training data. 
Given the source code, our approach enables dynamic trace collection for DNN.
We propose two techniques. The first technique, inspired by~\cite{gopinath2019symbolic2},
translates the code into an intermediate form which we call {\em imperative representation of the DNN}.
The purpose of the imperative representation is to make certain (ensure) that internal states of the DNN are observable, thus our method uses a 
{\em white box approach}.
This conversion to an imperative representation allows us to insert probes that
enable dynamic analysis over the traces produced by the DNN while it is 
being trained on the training data. 
The second technique uses a novel callback mechanism to insert probes that
also achieve the same purpose. 
We then conduct dynamic analysis over the traces to identify the faulty layer 
or hyperparameter that causes the error.
We also propose an algorithm to identify root causes by capturing any numerical 
error and monitoring the model during training and finding the relevance of 
every layer/parameter on the DNN outcome. 

We have implemented our approach as an extension of the widely used \keras library
for deep learning. 
To evaluate, we have collected a benchmark containing 40 buggy models and 
patches that contain real errors in deep learning application from  
\sof, and \gh.
Our benchmark can be used to evaluate automated debugging tools 
and repair techniques. 
We compare our approach with three built-in mechanisms for 
debugging in the \keras library, the state-of-the-art in DNN libraries.
These mechanisms were {\em TerminateOnNaN()}, {\em EarlyStopping ('loss')}, 
and {\em EarlyStopping ('accuracy')}.
We have evaluated our approach using this DNN bug-and-patch benchmark, and 
the result shows that our approach is much more effective than the existing 
debugging approach used in the state-of-the-practice \keras library. 
For 34/40 cases, our approach was able to detect faults whereas the 
best debugging approach provided by \keras detected 32/40 faults.
Our approach was able to localize 21/40 bugs whereas
\keras did not localize any faults.
In summary, this paper makes the following contributions:
\begin{itemize}
	\item We propose the {\em first fault localization approach for DNNs}
	including {\em callback and translation mechanisms for collecting dynamic traces} and a {\em localization algorithm}.

	\item We have built a {\em DNN bug-and-patch benchmark} with 40 different types of 
	buggy models from \sof and \gh.  
	This benchmark serves as the ground truth to evaluate our approach. 
	We also hope it can serve other researchers to validate their debugging and repair tools.
	This benchmark is available from \gh~\cite{myRepo}.
	\item Our results show that our approach can effectively and efficiently 
	identify 34 out of 40 buggy model and determine the root causes 
	for 21 out of 40 buggy model.
	
\end{itemize}

Outline:
\S\ref{sec:background} motivates our approach. 
\S\ref{sec:Approach} presents our dynamic trace collection, faulty detection and localization algorithms.
\S\ref{sec:EVALUATION} presents evaluation. 
\S\ref{sec:THREATSTOVALIDITY} discusses the threats to validity, 
\S\ref{sec:relatedwork} discusses related works and \S\ref{sec:conclusion} concludes.

\input{fig-overview}

%% file: fig-overview.tex
\begin{figure*}[h!t]
	\centering
	\includegraphics[width=1\linewidth,trim={0cm 0.2cm 0cm 0.2cm},clip]{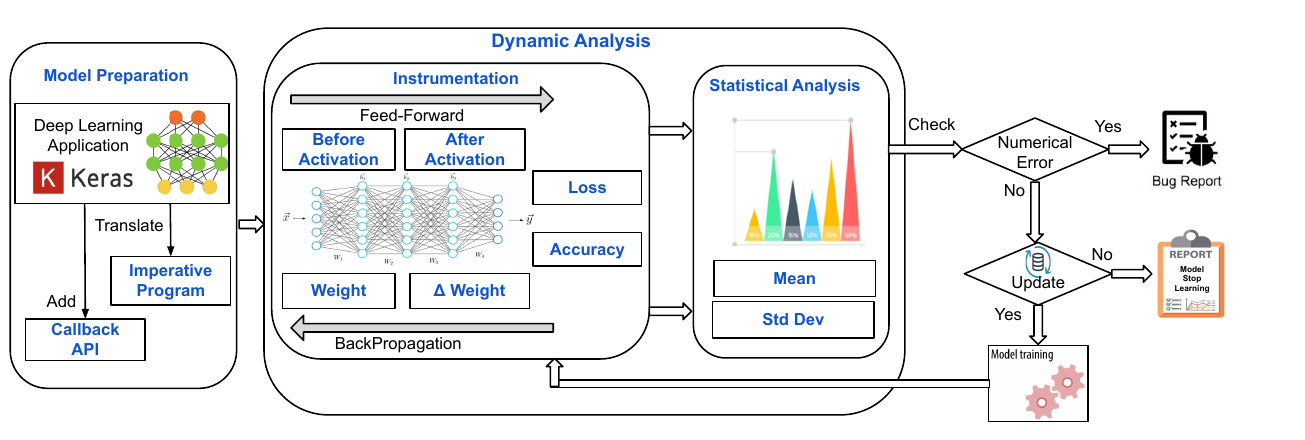}
	\caption{ An Overview of DeepLocalize. Left component shows two alternatives (callback and translation) for preparing models to collect dynamic traces. Middle component collects and analyzes these traces. Right component detects/localizes bugs.}
	\label{Process}
\end{figure*}

%% file: motivation.tex
\section{A Motivating Example}
\label{sec:background}

Listing~\ref{motivation-example} shows a simple example from \sof~\cite{bug1}
to motivate our work.  
In this example, the developer is constructing a sequential model at line 1,
adding a dense input layer at lines 2-3,
adding a dense hidden layer at lines 4-5,
compiling the model to convert it to a graphical form on line 6, and 
training this model on line 7. 
A dense layer is a layer in a DNN where each neuron is connected to neurons in the next layer.
This DNN did not learn during training and in the post, the 
developer asked why the model achieved low accuracy. 
This DNN has a two problems. 
First, it handles a classification problem, and thus \texttt{categorical\_crossentropy} 
should be used as a loss function at line~6 instead of \texttt{mean\_squared\_error}. 
Second, the user has not added activation functions for the 
first two layers at lines 2 and 4.

\begin{lstlisting}[basicstyle=\fontsize{9}{9}\ttfamily, language = python, caption={DNN is not learning \cite{bug1}},label=motivation-example]
model = Sequential()
layer1 = Dense(30, input_shape=input_shape, kernel_initializer=RandomNormal(stddev=1), bias_initializer=RandomNormal(stddev=1))
model.add(layer1)
layer2 = Dense(10, kernel_initializer=RandomNormal(stddev=1), bias_initializer=RandomNormal(stddev=1))
model.add(layer2)
model.compile(optimizer=SGD(lr=3.0), loss='mean_squared_error', metrics=['accuracy'])
model.fit(x_train, y_train, batch_size=10, epochs=30, verbose=2)
\end{lstlisting}

\keras provides a set of callback methods \cite{kerasCallBack} to give the 
developers information about internal status of the training process.  
Specifically, we can use \texttt{TerminateOnNaN()} to monitor the loss and 
terminate the training when the loss becomes \texttt{NaN}. 
We can use \texttt{EarlyStopping()} to monitor the loss or accuracy and stop 
if there is no improvement. 
We can pass a callback method as a parameter to the \texttt{fit()} method. 
For Listing~\ref{motivation-example}, when using
\texttt{TerminateOnNaN()}, \texttt{EarlyStopping('loss')}, 
\texttt{EarlyStopping('accuracy')}, and the union of the three \keras methods, the training was terminated after 
1.20, 12.21, 34.90 and  1.16 seconds respectively. 
Once the training is stopped, \keras prints the epoch and the iteration number. 
Unfortunately,  such information cannot answer the developer's question and 
indicate which layer or hyperparameter prevented the model from learning.


 
Our approach reports that the program has a bug after 0.14 seconds using our tool and 2.14 seconds using our callback function. 
In addition, we report that the bug is located in the back propagation stage of layer 2 
at line~4.  
Therefore, the message gives the developer hint that the issue in the parameter 
of layer 2, since the description of the message indicates the stage of the problem, 
the developer can quickly determine previous calculation that causes the problem, 
which is the loss function in our example.  
Compared to the existing methods in \keras, our fault localization uses less time 
and provides a report stating the layer that the bug is located at.

%% file: approach.tex
\section{Approach}
\label{sec:Approach}

\subsection{An Overview}

\fignref{Process} provides an overview of our work. 
In the first step, we prepare the DNN model to collect dynamic traces. 
We propose two mechanism for this. 
Our first approach translates the deep learning program into an imperative program~\cite{gopinath2019symbolic1, gopinath2019symbolic2}.
Probes are inserted in this imperative program to capture and save model variables 
such as weights, learning rate gradients during training.
Our second approach uses a callback mechanism, and passes a specialized callback
method as a parameter during model training (to the \texttt{fit()} method).
This custom callback function allows the developers to capture and save model variables.
We record the key values during both the feed-forward and backward propagation phases. 
During the training, an online statistical analysis is performed to compare 
the status of the program with the error conditions we defined. 
Finally, we report if the program contains a bug, and in which layer and 
phases the bug exists that prevented learning.

\subsection{Model Preparation}~\label{sec:translation}


\begin{table*}[htbp]
	\centering
	\Large
	\caption{Translation from \keras Code to an Imperative program}
	\scalebox{0.45}{
	\begin{tabular}{|r|p{40.5em}|p{36.5em}|}
		\hline
		\multicolumn{1}{|l|}{No} & \multicolumn{1}{c|}{\keras Code} & \multicolumn{1}{c|}{Imperative Program} \\
		\hline
		1     &     batch\_size = 1 &     batch\_size = 1 \\
		\hline
		2     &     nb\_epoch = 3 &     nb\_epoch = 3 \\
		\hline
		3     & \multicolumn{1}{r|}{\cellcolor[rgb]{ .98,  .984,  .988}\textcolor[rgb]{ .141,  .161,  .18}{}} & \cellcolor[rgb]{ .902,  1,  .929} + lr = 0.001 \\
		\hline
		4     &     model = Sequential() &     myModel = mySequential() \\
		\hline
		5     & \cellcolor[rgb]{ 1,  .933,  .941} - model.add(Dense(units=50, activation='relu', input\_dim=128)) & \cellcolor[rgb]{ .902,  1,  .929} + myModel.insert(myDense(num\_inputs=128, num\_outputs=50, lr\_rate=lr, name='FC1')) \\
		\hline
		6     & \multicolumn{1}{r|}{\cellcolor[rgb]{ .98,  .984,  .988}\textcolor[rgb]{ .141,  .161,  .18}{}} & \cellcolor[rgb]{ .902,  1,  .929} + myModel.insert(ReLu()) \\
		\hline
		7     &     model.add(Dropout(0.2)) &     myModel.insert(myDropout(0.2)) \\
		\hline
		8     & \cellcolor[rgb]{ 1,  .933,  .941} - model.add(Dense(units=50, activation='relu')) & \cellcolor[rgb]{ .902,  1,  .929} + myModel.insert(myDense(num\_inputs=50, num\_outputs=50, lr\_rate=lr, name='FC2')) \\
		\hline
		9     & \multicolumn{1}{r|}{\cellcolor[rgb]{ .98,  .984,  .988}\textcolor[rgb]{ .141,  .161,  .18}{}} & \cellcolor[rgb]{ .902,  1,  .929} + myModel.insert(ReLu()) \\
		\hline
		10    &     model.add(Dropout(0.2)) &     myModel.insert(myDropout(0.2)) \\
		\hline
		11    & \cellcolor[rgb]{ 1,  .933,  .941} - model.add(Dense(1,activation ='softmax')) & \cellcolor[rgb]{ .902,  1,  .929} + myModel.insert(myDense(num\_inputs =50, num\_outputs =1, lr\_rate =lr, name='FC3')) \\
		\hline
		12    & & \cellcolor[rgb]{ .902,  1,  .929}\textcolor[rgb]{ .141,  .161,  .18}{ + myModel.insert(Softmax())} \\
		\hline
		13    &  \cellcolor[rgb]{ 1,  .933,  .941}\textcolor[rgb]{ .141,  .161,  .18}{ - Adam = optimizers.Adam()} & \\
		\hline
		14    & \cellcolor[rgb]{ 1,  .933,  .941} - model.compile(loss ='binary\_crossentropy', optimizer =Adam, metrics=['acc']) & \cellcolor[rgb]{ .902,  1,  .929} + myModel.myCompile(loss='binary\_crossentropy', optimizer='Adam', metrics=['acc']) \\
		\hline
		15    & \cellcolor[rgb]{ 1,  .933,  .941} - model.fit(X\_train, Y\_train, batch\_size=batch\_size, nb\_epoch=3, verbose=1, validation\_data=(X\_test, Y\_test)) & \cellcolor[rgb]{ .902,  1,  .929} + myModel.fit\_instrument(X\_train, Y\_train, batch\_size=batch\_size, epoch=3) \\
		\hline
		16    & \cellcolor[rgb]{ 1,  .933,  .941} - model.evaluate(X\_test, Y\_test, verbose =1)  & \cellcolor[rgb]{ .902,  1,  .929} + myModel.myEvaluate(X\_test, Y\_test, 200)  \\
		\hline
	\end{tabular}%
	\label{tab:Translation}%
}
	\begin{tablenotes}
	\small
	\item This table is showing all of the changes to translate from \keras code to Our tool code.
	The color in each row indicates the change type: 
	\color{green}
	{Green + added a new line}, \color{red}
	{Red - removed an existing line}.
	\end{tablenotes}

\end{table*}%


 Directly analyzing a deep learning program, e.g. one shown in Listing~\ref{motivation-example} is hard, 
 as the DNN libraries provide blackbox APIs and it is hard to trace important values during training. 
 To use our approach, the developer either write extra code to instrument DNN training inside 
 \texttt{fit()} function or rewrite their code into an imperative form. 
 For our second approach, we identified a list of the \keras library APIs that are important for 
 training and implemented models/simplified versions of these API calls, following the 
 machine learning literature~\cite{sim2017improving, hertz2018introduction, ruder2016overview} 
 and the \keras documentation~\cite{kerasDoc}. 
 We inserted the probes to these library models so that the analysis can observe the 
 internal behaviors of DNNs during training. 

The callback-based approach of dynamic trace collection is implemented 
by overriding the \texttt{keras.callbacks.Callback} class.
Since our work is focusing on monitoring during the training phase, 
we override the method called (\texttt{on\_train\_batch\_end(self, batch, logs=None)}).
This overridden method invokes Algorithm~\ref{alg:fault-localize}
after each batch of training.
To use this method, the developer needs to pass this custom callback 
function as a parameter to the \texttt{fit()} function.

 
Second imperative approach, as shown in \tabref{tab:Translation} on the left, to build a training model, a DNN program typically starts 
 with creating a sequential model (line 4), then add all layers (lines 5--11) and optimizations (line 13), 
 and finally call {\tt compile} and {\tt fit} at lines 14--15. 
 On the right, we show the imperative programs using our library models. 
 First, lines 1, 2, 7 and 10 are not changed. 
 Second, lines 5, 8, and 11 show the conversions for the Dense layer. 
 In our code, we added a name for the layers, e.g., see name = "FC1" so that we can 
 report in which layer the fault is located. 
 We inserted instrumentation in the fit function (line~15) to observe the model variables. 
 
 
Currently, our translation tool supports the \texttt{Dense}, \texttt{Dropout}, \texttt{Maxpooling}, \texttt{Convolution}, 
\texttt{BatchNormalization}, and \texttt{Padding} layers. 
Also, it supports popular optimization methods, losses, and activation functions. 
The translation from a DNN program to the program that uses our library models is 
currently done manually. 
The \keras library is being rapidly evolved resulting in a large number of releases \cite{islam20repairing}. 
Due to the library versioning and the frequently changing API signatures, 
it is hard for our tool to remain compatible with the \keras library.

\subsection{Instrumentation}
\label{sec:Instrumentation}
 
The training of DNN has two phases: feedforward and backpropagation.
In the feedforward stage, we observe and monitor: (1) the training data including 
both input and label; (2) the results after applying forward function; 
(3) the results after applying the activation function; 
(4) the loss value and (5) the accuracy. 
All values are collected in each iteration during training. 

The second stage is the backpropagation, and it is used to adjust the weight 
based on the errors obtained from the feedforward stage. 
Backpropagation uses the gradient descent method to update the weights and 
minimize the errors. 
It is started from the output layer, and the result of the output layer is reused 
to compute the gradient for the previous layer until it reaches the input layer. 
Different optimizations can be applied during backpropagation. 
In the backpropagation stage, we can observe and monitor: (1) the update 
of weights, (2) the update of bias values, and $\Delta$ weights from applying 
the gradient descent for each layer.

The instrumentation is inserted in the \texttt{fit()} function we implemented for modeling the \keras \texttt{fit()} function. It is executed automatically when the DNN program runs during training.





\subsection{Statistical Analysis to Detect Suspect Behavior}
\label{sec:NumericalErrorDetection}

Next, we discuss our approach for statistical analysis to detect suspect behavior 
of the DNN during training. 
We analyze three variables: the learning rate, the input data, and activation/loss functions.



\subsubsection{Incorrect learning rate}

Backpropagation is important to fine-tune the weights based on the loss value obtained from the loss function. The learning rate has an effect on the weight updates during the backpropagation process.  During the backpropagation process, the learning library computes gradient descent iteratively. Our key insight is that the mean and standard deviation of the weights in the correct model are continuously changing during the training process. In contrast, the mean and standard deviation of the weights in the buggy model are constant. If there is a problem in the learning rate, we can detect it from the output of the gradient for the output layer. \fignref{fig:2figsA} and \fignref{fig:2figsB} show an example of this behavior. In \fignref{fig:2figsA} the weight varies as it should in a correct model, whereas in \fignref{fig:2figsB} the weight is constant indicating a potential bug. To utilize this insight, we compute the mean and the standard deviation for the output of the gradient as well as the weight parameter at each layer. 


\subsubsection{Incorrect input data} 
In some cases, the training data are not properly normalized. For example, in the MNIST model, the pixel should be in the range [-1, 1] and not [0, 255]. Also,  training data may have \texttt{NaN} value, and developers forget to invoke the assert function to check if there is an \texttt{NaN} or not. In the forward stage, we retrieve the output before/after applying the activation function for each year. We then compute the mean and standard deviation for the output at each layer. We will check if the output of the first layer after/before applying activation has numerical error such as {\tt NaN} or {\tt Inf}. Our second approach for detecting this kind of error is to calculate how frequent the mean of the output for the first layer is equal to zero. Once we observe abnormal values, we will report in which layer and whether it is before or after activation function that the error occurs.


\subsubsection{Incorrect activation/loss functions} 
After finishing the forward stage, we compute loss and accuracy. There are two indicators that the model has a problem.  First, if one of the two metrics has a numerical error like {\tt inf} or {\tt NaN}. Second, the loss starts increasing instead, or the accuracy starts decreasing after certain iterations.

\begin{figure}[h!t]
	\centering
	\includegraphics[width=3.4in,trim={1.5cm 9cm 1cm 9.5cm},clip]{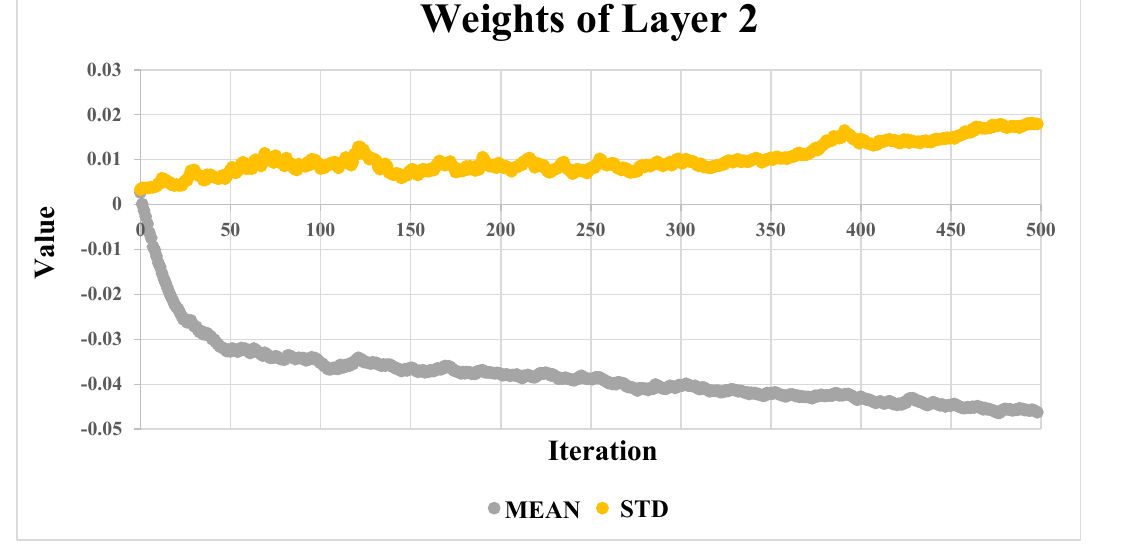}
	\caption{Weight of correct model}
	\label{fig:2figsA}
\end{figure}
\begin{figure}[!t]
	\centering
	\includegraphics[width=3.4in,trim={1.5cm 5cm 1cm 4.5cm},clip]{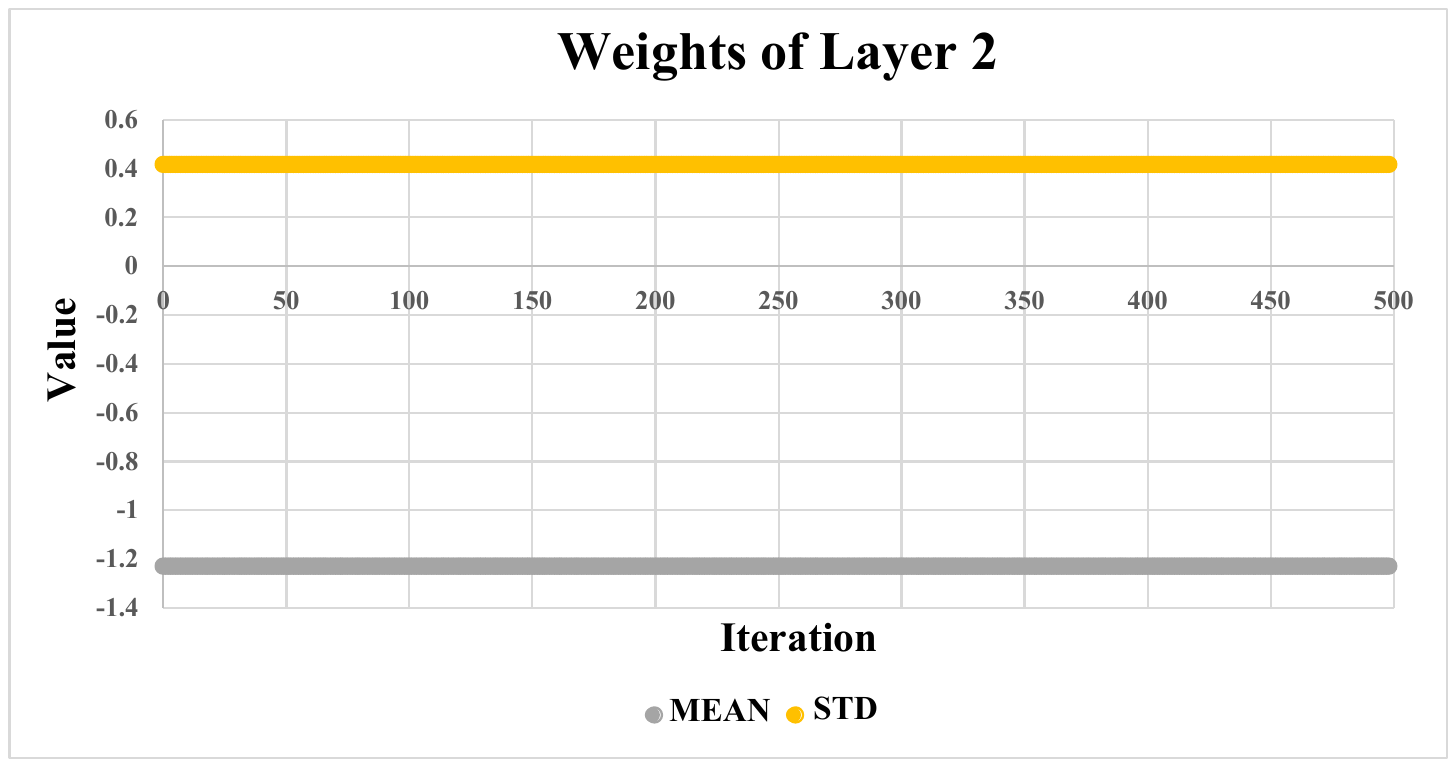}
	\caption{Weight of buggy model }
	\label{fig:2figsB}
\end{figure}

\subsection{DNN Fault Localization Algorithm}
\label{sec:Algorithem}

Algorithm~\ref{alg:fault-localize} presents our DNN fault localization algorithm.
At its core, it augments the DNN learning algorithm with analysis and error 
checking inserted during learning. 
It serves three purposes: (1) determining whether the deep learning program 
contains a bug; 
(2) reporting the fault location, in which layer and which phases (feed 
forward and backward propagation), the deep learning program has a bug; 
(3) reporting failure information, in which iteration, the learning is stopped. 
It takes as input the training data set (including input and labeling), 
the translated imperative program, as well as the DNN parameters (the batch 
size and the epoch). 
If the bug is found, the output is a message including the fault location 
and failure information for the bug.


At line 1, we define two lists to store the values of loss and accuracy 
in each iteration. 
Line 2 represents how many iterations we make through the whole training dataset. 
During the training process, the training dataset is divided into several 
smaller batches. 
For example, if the model has 2000 training examples divided into 500 batches, 
then the model needs 4 iterations to complete one epoch. 
Line 3 shows the division of the batch size. 
Lines 4-29 run for each batch in the training dataset. 
At lines 5--11, the algorithm performs dynamic analysis on the forward stage, 
and at lines 24--29, it performs analysis on the backward stage. 
In our callback function, the  override method \texttt{on\_train\_batch\_end(self, batch, logs=None)} 
will execute at the end of each batch, and performs dynamic analysis on the
forward/backward stage, after retrieving the value of each layer before/after 
activation function, loss, accuracy, updating weight and gradients. 

\subsubsection{Feedforward phase} 
At line 6, the algorithm computes the output of a feed-forward layer before applying the activation function. At line 8, we compute the output after applying the activation function. Then, we invoke ANA() procedure to determine if the output contains a numerical error. As shown in  \tabref{tab:freq}, the message EBA indicates Error Before Activation, EAA indicates Error After Activation, and L represents the faulty layer number. At line 12, we compute the loss, and determine if there is any numerical error at line 13.  As shown int \tabref{tab:freq}, the ELF indicates Error in Loss Function. If the error is not detected here, we save the loss value for each iteration at line 14. At line 15, we compute the accuracy and check if there is a numerical error from accuracy at Lines 16-18. If the error is not detected here, we save the accuracy value during training at Line 19. 
At lines 20-22, the algorithm checks if loss is not decreasing and accuracy is not increasing value for a long time. In both cases, The algorithm computes the slope to compare the loss/accuracy for current step with the loss/accuracy at a step which is at least num\_steps behind the current step. In this case, the algorithm reports a message MDL to indicate that the model does not learn. Otherwise, the training continues.

\subsubsection{Backpropagation phase} 
During this stage, the algorithm collects the weight and $\Delta$ weight for each layer in each iteration. At line 25, the Weight and $\Delta$ Weight are the output of a back-propagation. At line 26, the algorithm invokes the ANA() procedure and pass $\Delta$ weight to check if there is any numerical error. The algorithm will print error message if there is any error in the $\Delta$ weight and determines which layer causes this issue (Line 26). In the same way, at line 27, the algorithm can determine if there is an issue in the weight in each layer by invoke the ANA() procedure. If the procedure decide that there is an issue in the weight, then the algorithm will return message to indicate there is bug.

Finally, if there are no issues in the model, the algorithm will terminate after finishing training at line 32 and print this message Correct Model (CM).

\begin{table}
	\caption{Abbreviation of Crash Statements}
	\label{tab:freq}
	\centering
	\begin{tabular}{ccl}
		\hline
		No&Statement&Abbreviation\\
		\hline
		1 & Error Before Activation & EBA\\
		2 & Error After Activation & EAA\\
		3 & Error in Loss Function, & ELF\\
		4 & Error in Accuracy Function & EAF\\
		5 & Error Backward in Weight & EBW\\
		6 & Error Backward in $\Delta$ Weight & EBDW\\
		7 & Model Does not Learn  & MDL\\
		8 & Correct Model  & CM\\
		\hline
	\end{tabular}
\end{table}


{\tt ANA()} is invoked at lines 7, 9, 26 and 27 to determine if the error occurs based on the current values obtained from the instrumentation.  When the DNN does not learn, there can be the following symptoms: (1) the update for weight, $\Delta$ weight are incorrect, (2) the loss or accuracy is not measured on the correct scale, and (3) the loss does not decease, and the accuracy does not increase after the number of iterations. The check is conducted in {\tt ANA()}.



This procedure takes three parameters: input value, layer number and location. Since {\tt ANA()} is called at different locations in the code, the location tracks whether the value comes from feed-forward (FW), backward (BW) propagation, or weight (WT). Line 2 defines a set of lists to store the mean value from each location for each layer. Line 3 computes the mean value for the input. At line 4, the procedure will check if the mean reports \texttt{NaN} or \texttt{inf}. Also, the procedure will check if the mean equal to zero  at line 5; if yes, then we compute how frequent zero occurs for all values for each layer. If the number of zeroes is greater than a threshold (line 7), the error is detected and the procedure will return true.

At line 11, we store the mean value in the list. Finally, the procedure will return the last N element from the list as slice  to check if the mean value for last N iterations is changed or not. From \fignref{fig:2figsA} and \fignref{fig:2figsB}, we observe the model continues to learn if the mean value is changing in each iteration. This procedure returns true if there is a numerical error; otherwise, it return false. 


\IncMargin{1.0em}
\begin{algorithm}
 \caption{DNN Fault Localization}
 \label{alg:fault-localize}
 \small
 \DontPrintSemicolon
  \SetKwData{Left}{left}
  \SetKwData{Up}{up}
  \SetKwFunction{Forward}{Activation}
  \SetKwInOut{Input}{input}
  \SetKwInOut{Output}{output}

\Indm\Indmm
  \Input{Training data (input, label), batchsize, epochs, imperative program}
  \Output{A message regarding fault location and failure information}
\Indp\Indpp
  \BlankLine
  $LossList \leftarrow []$; $AccuList \leftarrow []$\;
  \For{$e\leftarrow 0$ \KwTo $epoches$}{ 
    \For{$i\leftarrow 0$ \KwTo $Length(input)$ \KwStep $batchsize$ }{
        $X \leftarrow input[i]$; 
		$Y \leftarrow label[i]$\;
		\For{$L\leftarrow 0$ \KwTo $Length(Layers)$  }{
		    $V_1$ $\leftarrow$ $Layer[L].Forward(X)$\;
		    \lIf{$ANA(V_1, L, FW)$}
                {
                \Return EBA, L
                }
            $V_2= Layer[L].Activation(V_1)$\;
		    \lIf{$ANA(V_2, L, AF)$}
                {
                \Return EAA, L
                }
            $X \leftarrow V_2$\label{algo2:7}
            
        }
        $Loss  \leftarrow ComputeLoss(V_2, Y)$\;
        \lIf{$Loss$ is equal to NaN OR $inf$}
        {
		\Return ELF\label{algo2:44}
		}
		$LossList.append(Loss)$\;
		$Accuracy \leftarrow ComputeAccuracy(V_2, Y)$\;
		\If{$Accuracy$ is equal to NaN OR inf OR $0$  }
		{
		 \Return EAF \label{algo2:44}
	    }
	    
		$AccuList.append(Accuracy)$\;
		\If{$NOTDecreasing(AccuList)$ \&\&  $NOTIncreasing(LossList)$}
		{
		    \Return MDL \label{algo2:44}
		}
		
        $dy \leftarrow Y$\;
			\For{ $L \leftarrow Length(Layers)$ \KwTo $0$} {
				$V_3, W[L] \leftarrow Layer[L].Backward(dy)$\; 
				\lIf{$ANA(V_3, L, BW)$  }
				{
				    \Return EBDW, L\label{algo2:44}
				}
				\lIf{$ANA(W[L], L, WT)$  }
				{
    				 \Return EBW, L\label{algo2:44}
				}
				$dy \leftarrow V_3$\;
			}
        }
    }
    \Return CM\;
    \BlankLine
    \Indm\Indmm
\Indp\Indpp
\end{algorithm}
\DecMargin{1.0em}

\begin{algorithm}
  \DontPrintSemicolon
  \SetKwFunction{FANA}{ANA}
  \SetKwProg{Fn}{Function}{:}{}
  \Fn{\FANA{$input$, $LayerNo$, $Location$}}{
        $meanList[LayerNo, Location] \leftarrow []$\;
		$meanValue \leftarrow math.mean(input)$\; 
		\lIf{$meanValue$ is equal NaN OR inf }
		{
		\Return True
		}
		\If{$meanValue$ == $0$}
		{
		$FreqZero[LayerNo, Location]$    $+= 1$\;
		\If{$FreqZero[LayerNo, Location]$ $\geq$ $threshold$ }
    		{
    		    \Return True\;
    		}
		}
		$meanList[LayerNo, Location].append(meanValue)$\;
		$ slicing  \leftarrow meanList[LayerNo, Location][-N:]$\;
		\If{$All(elem == slicing[0]$ $for$ elem in slicing$)$ } 
		{
		\Return True \;
		}	    
		\Return False\;
  }
\end{algorithm}

%% file: evaluation.tex
\section{Evaluation}
\label{sec:EVALUATION}

\begin{table}[htbp]
	\centering
	\caption{\keras result VS Imperative result}
	\footnotesize
	\begin{adjustbox}{width=\columnwidth,center}
		\begin{tabular}{|r|r|r|r|r|r|r|r|r|}
			\hline
			\multicolumn{1}{|c|}{\multirow{3}[6]{*}{\textbf{Post \#}}} & \multicolumn{1}{c|}{\multirow{3}[6]{*}{\textbf{Epoch}}} & \multicolumn{1}{c|}{\multirow{3}[6]{*}{\textbf{Iteration}}} & \multicolumn{3}{c|}{\textbf{\keras}} & \multicolumn{3}{c|}{\textbf{Our Tool}} \\
			\cline{4-9}          &       &       & \multicolumn{1}{c|}{\textbf{Runtime}} & \multicolumn{2}{c|}{\textbf{Result}} & \multicolumn{1}{c|}{\textbf{Runtime}} & \multicolumn{2}{c|}{\textbf{Result }} \\
			\cline{5-6}\cline{8-9}          &       &       & \multicolumn{1}{c|}{\textbf{[sec]}} & \multicolumn{1}{c|}{\textbf{Loss}} & \multicolumn{1}{c|}{\textbf{Accuracy }} & \multicolumn{1}{c|}{\textbf{[sec]}} & \multicolumn{1}{c|}{\textbf{Loss}} & \multicolumn{1}{c|}{\textbf{Accuracy }} \\
			\hline
			\rowcolor[rgb]{ .859,  .859,  .859} 48385830 & 30    & 60000 & 489.67 & NaN   & 0.10  & 1429.60 & NaN & 0.10 \\
			\hline
			\textbf{31556268} & 1000  & 1     & 5.58  & 0.50  & 0.50  & 20.99 & 0.50  & 0.50 \\
			\hline
			\rowcolor[rgb]{ .859,  .859,  .859} 50306988 & 5     & 200   & 1.41  & 0.71  & 0.50  & 0.17  & 0.69  & 0.50 \\
			\hline
			\textbf{48251943} & 17    & 500   & 5.40  & 0.37  &  --   & 2.10  & 0.30  &  -- \\
			\hline
			\rowcolor[rgb]{ .859,  .859,  .859} 38648195 & 20    & 48000 & 123.21 & 0.23  & 0.75  & 385.40 & NaN   & 0.65 \\
			\hline
			\textbf{33969059} & 20    & 10000 & 129.85 & 349713063.30 & -- & 1272.30 & 260179130.95 & -- \\
			\hline
			\rowcolor[rgb]{ .859,  .859,  .859} 55328966 & 10    & 49999 & 104.95 & 2.30  & 0.10  & 8025.54 & 2.29  & 0.11 \\
			\hline
			\textbf{34311586} & 20    & 6     & 1.23  & 0.67  &  --   & 0.11  & 0.50  &  -- \\
			\hline
			\rowcolor[rgb]{ .859,  .859,  .859} 31880720 & 3     & 20000 & 181.30 & 7.67  & 0.50  & 529.29 & 16.12 & 0.50 \\
			\hline
			\textbf{39525358} & 150   & 13    & 3.15  & 0.67  & 0.62  & 5.16  & 0.67  & 0.61 \\
			\hline
			\rowcolor[rgb]{ .859,  .859,  .859} 48934338 & 1000  & 50    & 10.17 & 1358.22 & 0.00  & 28.58 & 787.75 & 0.00 \\
			\hline
			\textbf{47724077} & 50    & 32561 & 39.26 & 49.70 & 0.68  & 93.46 & 3.38  & 0.70 \\
			\hline
			\rowcolor[rgb]{ .859,  .859,  .859} 59278771 & 200   & 135   & 18.91 & 1.10  & 0.35  & 14.86 & 1.06  & 0.34 \\
			\hline
			\textbf{41372874} & 20    & 239   & 3.19  & 0.15  & 1.00  & 6.12  & 0.00  & 1.00 \\
			\hline
			\rowcolor[rgb]{ .859,  .859,  .859} 44066044 & 100   & 21    & 3.86  & 9.55  & 0.38  & 0.89  & 9.55  & 0.00 \\
			\hline
			\textbf{51930566} & 50    & 75    & 4.04  & 0.51  & 0.67  & 1.87  & 0.48  & 0.64 \\
			\hline
			\rowcolor[rgb]{ .859,  .859,  .859} 45442843 & 100   & 200   & 2.74  & 0.50  & 0.50  & 23.34 & 0.51  & 0.50 \\
			\hline
			\textbf{31627380} & 10    & 712   & 5.75  & 1.29  & 0.59  & 8.31  & 1.46  & 0.63 \\
			\hline
			\rowcolor[rgb]{ .859,  .859,  .859} 58609115 & 10    & 442   & 2.63  & 26169.69 & 0.09  & 13.10 & 4.05  & 0.02 \\
			\hline
			\textbf{50481178} & 50    & 200   & 11.82 & 0.02  & 0.99  & 6.13  & 0.01  & 0.98 \\
			\hline
		\end{tabular}%
		\label{tab:ImperativeVSKeras}%
	\end{adjustbox}
\end{table}%


In this section, we aim to answer the following research questions: 
\begin{itemize}
	\item RQ1 (Validation): Can our technique find bugs in deep learning programs effectively?
    \item RQ2 (Comparison): How effectively and how fast can our technique localize the faults compared to existing methodology in the \keras library?
    \item RQ3 (Limitation): In which cases does our technique fail to report the bug and localize the faults?

\end{itemize}

\subsection{Experimental setup}


\subsubsection{Implementation} 
To perform our experiments and evaluation, 
we implemented our techniques using Python and \keras. 
Our translation-based tool supports the \texttt{Dense}, \texttt{Dropout}, \texttt{Maxpooling}, 
\texttt{Convolution}, \texttt{BatchNormalization}, and \texttt{Padding} layers. 
Also, it supports popular optimization methods, losses, and activation functions. 
We followed the machine learning references \cite{sim2017improving, hertz2018introduction, ruder2016overview} and used the \keras documentation 
to implement simplified and instrumented versions of \texttt{ compile()}, 
{\tt sequential()}  and \texttt{fit()} function. 
Our callback-based tool supports all of the layers and optimization
methods supported by \keras.

We set $threshold$ = 1/4 * No. Iteration and N= 50 at line 6 and 12 
in \texttt{ANA()} function respectively for both our tool and 
callback function. 
Based on our empirical experience, these settings helped best in 
detecting bugs during training.



\subsubsection{Benchmark construction} 
We collected buggy models from \sof posts 
 and \gh commits to construct the benchmark. 
 In \sof, we select the posts that have a score $\geq$ 5 
 and contains the buggy \keras code. 
 We used keywords \texttt{"error"},  \texttt{"bug"}
 and synonyms to search for posts.
%
%
In the second step, we manually reviewed the retrieved posts and removed 
all the posts that have partial code.
%
In the third step, we applied a second filter by checking if the post 
provided the training data or used the existing known training data. 
In the last step, we studied all the answers corresponding to the 
post ids in the \sof, using the methodology in 
a prior work~\cite{islam20repairing} that studied the bug fix patterns 
for deep learning model. 
We take into consideration the acceptable quality metric and choose 
the answer that has the highest score. 
We analyzed each Q\&A and derived for each post the fault location 
and a patch to fix the bug. 
In total, we obtained 30 posts.

\begin{figure}
	\centering
	\includegraphics[width=3.4in,trim={1.5cm 9cm 1cm 8.5cm},clip]{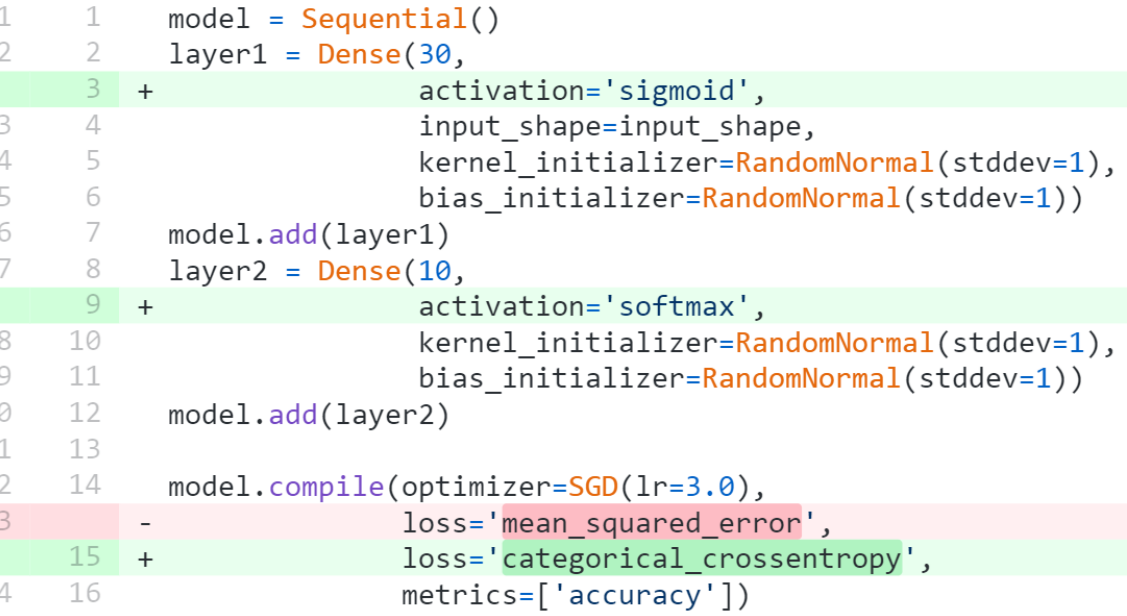}
	\caption{(Patch/Fix) the changing to fix a DNN bug~\cite{bug1}}
	\label{fig:patch}
\end{figure}

We also mine the \gh commits to collect buggy models. 
The process consists of three steps. 
First, we search for all \keras repository. 
After that, we mine all the commits whose title contains keywords used 
in the \sof mining process described above. 
Then, we take only the last version fix, and manually check if 
the commits are related to structure bug or not, from these 
commits, we derived the fault location and a patch to fix the bug. 
In the last step, we checked if the repository provided the training data. 
As a result, we randomly select 11 executable  programs with training dataset. 
For instance,  \fignref{fig:patch} shows the patch we derived for 
the  \sof post \# \cite{bug1}. Table~\ref{tab:Benchmarks} and \ref{tab:BenchmarksGitHub} 
present our benchmark together with  total number of parameters, 
the line of code, its loss and accuracy before and after fix. 
Our tool and benchmarks are available for download \cite{myRepo}.







\tabref{tab:ImperativeVSKeras} shows a comparison between the original \keras code and our imperative code in terms of accuracy, losses and runtime. Over all benchmarks, our imperative code has comparable accuracies and losseses compared to \keras. The small deviation is due to additional optimizations applied in \keras code. In terms of runtime, our imperative code takes around 5 times more execution time compared to \keras during training and testing phases. The reason behind that is the extra work done by our imperative code to identify bugs and their root causes. Our code is also able to terminate early when a bug is found and later in our results, we show that we can detect bugs faster than \keras callback methods.

All the experiments are run on a computer with Intel(R) Core (TM) i7-6500U, 2.5 GHz processor, and 16 GB of RAM running the 64-bit Windows 8.1 operating system.



\begin{table}[htbp]
	\centering
	\caption{The Benchmark, we show (Post \#) post id from \sof, parameters \#, line of code, (Result Before Fix) the loss and accuracy before applying patch, and (Result After Fix) the loss and accuracy value after applying patch}
	\scalebox{0.75}{
	\begin{tabular}{|r|r|r|r|r|r|r|}
		\hline
	
				\multirow{2}[4]{*}{\textbf{Post \#}} & \multicolumn{1}{c|}{\multirow{2}[4]{*}{\textbf{Total params}}} & \multirow{2}[4]{*}{\textbf{LOC}} & \multicolumn{2}{p{8.11em}|}{\textbf{Result Before Fix}} & \multicolumn{2}{p{8.11em}|}{\textbf{Result After Fix}} \\
				\cline{4-7}          &       &       & \textbf{Loss} & \textbf{Accuracy} & \textbf{Loss} & \textbf{Accuracy} \\
				
		\hline
		\rowcolor[rgb]{ .859,  .859,  .859} 48385830 &23,860 & 60 & NaN   & 0.10  & 0.15  & 0.96 \\
		\hline
		44164749 & 6,931,610& 49& 0.38  & 0.89  & 0.20  & 0.92 \\
		\hline
		\rowcolor[rgb]{ .859,  .859,  .859} 31556268 &17 & 39& 0.25  & 0.50  & 0.00  & 1.00 \\
		\hline
		50306988 & 12& 45& 0.70  & 0.50  & 0.67  & 1.00 \\
		\hline
		\rowcolor[rgb]{ .859,  .859,  .859} 48251943 & 17& 30 & 0.37  & { --} & 0.63  & { --} \\
		\hline
		38648195 & 903 & 36& 0.22  & 0.41  &      1.3  & 0.65 \\
		\hline
		\rowcolor[rgb]{ .859,  .859,  .859} 33969059 &23 & 63& 72181941.16 & { --} & 9766327.81 & { --} \\
		\hline
		55328966 &655,200 & 52& 2.30  & 0.10  & 0.85  & 0.80 \\
		\hline
		\rowcolor[rgb]{ .859,  .859,  .859} 34311586 &19 & 29&  0.67  &{ --} & 0.17  & { --} \\
		\hline
		31880720 & 9,051 & 46 & 7.67  & 0.50  & 0.00  & 1.00 \\
		\hline
		\rowcolor[rgb]{ .859,  .859,  .859} 39525358 &91 &34 & 0.67  & 0.62  & 0.31  & 1.00 \\
		\hline
		39217567 & 216 & 43& 0.25  & 0.50  & 0.08  & 0.97 \\
		\hline
		\rowcolor[rgb]{ .859,  .859,  .859} 48934338 &1,661 & 37 & 1335.72 & { --} & 341.09 &{ --} \\
		\hline
		47724077 & 359& 46& 29.97 & 0.69  & 0.55  & 0.76 \\
		\hline
		\rowcolor[rgb]{ .859,  .859,  .859} 59325381 & 221,226 & 38 & 11023.33 & 0.10  & 0.03  & 0.99 \\
		\hline
		59278771 &35 & 43 & 1.10  & 0.35  & 0.39  & 0.83 \\
		\hline
		\rowcolor[rgb]{ .859,  .859,  .859} 52800582 &2,701 & 41 & 11112.00 & { --} & 0.00  & { --} \\
		\hline
		41372874 & 15& 39& 0.15  & 1.00  &     0.15  & 1.00  \\
		\hline
		\rowcolor[rgb]{ .859,  .859,  .859} 34673164 & 5,378& 50 & 0.13  & 0.78  & 0.43  & 0.89 \\
		\hline
		48221692 &16 & 28& 2311.60 & {   --} & 121.94 & { --} \\
		\hline
		\rowcolor[rgb]{ .859,  .859,  .859} 50079585 & 1,212,513& 75 & 9297.00 & 0.33  & 0.94  & 0.55 \\
		\hline
		45337371 & 3,221& 31 & 7.85  & { -- } & 0.69  & { --} \\
		\hline
		\rowcolor[rgb]{ .859,  .859,  .859} 44066044 & 313& 38 & 9.55  & 0.38  & 0.01  & 1.00 \\
		\hline
		51930566 & 35 & 52& 0.77  & 0.64  & 0.30  & 0.91 \\
		\hline
		\rowcolor[rgb]{ .859,  .859,  .859} 47352366 & 3,274,634& 50& NaN   & 0.06  & 0.10  & 0.98 \\
		\hline
		45442843 &4 & 41 & 0.50  & 0.50  & 0.65  & 0.83 \\
		\hline
		\rowcolor[rgb]{ .859,  .859,  .859} 48594888 & 4,873,738 & 60 & 1.45  & 0.48  & 0.86  & 0.70 \\
		\hline
		31627380 & 6,658 & 64 & 1.24  & 0.58  & 0.83  & 0.63 \\
		\hline
		\rowcolor[rgb]{ .859,  .859,  .859} 58609115 & 9,798 & 43 & 44539.79 & 0.10  & 0.07  & 0.99 \\
		\hline
		50481178 &4,241 & 43 & 0.02  & 0.99  & 0.02  & 1.00 \\
		\hline
	\end{tabular}}%
	\label{tab:Benchmarks}%
\end{table}%

\begin{table}[htbp]
	\centering
	\caption{The Benchmark, we show (GH $\mid$ Ref) the \gh repository reference the number of parameters, line of code (LOC), (Result Before Fix) the loss and accuracy value before applying patch, and (Result After Fix) the loss and accuracy after applying patch}
	\scalebox{0.85}
	{
	\begin{tabular}{|r|r|r|r|r|r|r|}
		\hline
	
				\multirow{2}[4]{*}{\textbf{GH $\mid$ Ref}} & \multicolumn{1}{c|}{\multirow{2}[4]{*}{\textbf{Total params}}} & \multirow{2}[4]{*}{\textbf{LOC}} & \multicolumn{2}{p{8.11em}|}{\textbf{Result Before Fix}} & \multicolumn{2}{p{8.11em}|}{\textbf{Result After Fix}} \\
				\cline{4-7}          &       &       & \textbf{Loss} & \textbf{Accuracy} & \textbf{Loss} & \textbf{Accuracy} \\
				
		\hline
		1 $\mid$ \cite{GitHub2} & 18989 & 79 & 0.39 &	0.90 &	0.38 &	0.90 \\
		\hline
		\rowcolor[rgb]{ .859,  .859,  .859} 2 $\mid$ \cite{GitHub3} & 367107 &	39  & 0.82 & 0.76 & 0.83 & 0.75 \\
		\hline
		3 $\mid$ \cite{GitHub5} & 468980 &	72 & 0.72 &	0.16 & 0.00 & 0.75 \\
		\hline
		\rowcolor[rgb]{ .859,  .859,  .859} 4 $\mid$ \cite{GitHub6} &34438 &	107  	 & 0.65 &	0.61 &	0.80 &	0.71\\
		\hline
		5 $\mid$ \cite{GitHub7} &169431	& 73  	 & 0.30 &	0.91 &	0.46 &	0.86\\
		\hline
		\rowcolor[rgb]{ .859,  .859,  .859} 6 $\mid$ \cite{GitHub11} & 221	& 16 	 & 0.48	 & 0.78 &	0.29 &	0.88\\
		\hline
		7 $\mid$ \cite{GitHub17} & 1941 &	915	 & 44.85 & --  & 44.75 & --\\
		\hline
		\rowcolor[rgb]{ .859,  .859,  .859} 8 $\mid$ \cite{GitHub22} &  341 &	183 & 0.02 & 0.99 &0.00 & 1.00\\
		\hline
		9 $\mid$ \cite{GitHub23} & 160305 &	40	 & 0.28 & 0.92 & 0.18 & 0.94 \\
		\hline
		\rowcolor[rgb]{ .859,  .859,  .859} 10 $\mid$ \cite{GitHub25} & 110627418	 & 67	 & 0.65	 & 0.63 & 	0.43 &	0.79 \\
		\hline
		11 $\mid$ \cite{GitHub27} &  9 &	30	 &1E-05 &	 --	 & 1E-05 &	 -- \\
		\hline
	\end{tabular}}%
	\label{tab:BenchmarksGitHub}%
\end{table}%


\subsection{Results and Analysis}



In \tabref{tab:Comparison} and \ref{tab:ComparisonGitHub}, the first column reports the \sof post \# with the corresponding link, and \gh repository  reference respectively. To compare our results with the results generated from the \keras methods, we listed the columns of Time, Epoch, Iteration, IB, and FL, representing how long the approach takes to find the bug, at which epoch and at which iteration the bug is reported, whether the approach identifies the bug correctly, and whether the approach successfully reports the fault location of the bug, respectively. Under {\it IB} and {\it FL}, \greencheck means that the approach is successful, 
 X means it fails, and -- is not able to identify or fault the bug.


The results show that our tool is able to identify bugs for 23 out of 29 buggy programs, and our callback function is able to identify bugs for 34 out of 40 buggy programs. In contrast, \texttt{TerminateOnNaN()}, \texttt{EarlyStopping('loss')}, \texttt{EarlyStopping('accuracy')}, and the
three \keras methods together are able to identify 2, 24, 28, and 32 out of 40 bugs respectively.
This is mainly because our technique used a wider spectrum of important runtime values for analysis compared to the other three methods, which only used one metric such as the loss, or accuracy. Also, \texttt{EarlyStopping('loss')} and \texttt{EarlyStopping('accuracy')} sometimes do not give a good indication that the model has a problem. For example, the deep neural network may be stuck at a local minimum at some point that has values that
are close to the global minimum \cite{kawaguchi2016deep}.  

 We failed to detect bugs for 7 out of 30 models using our tool. Our tool is not able to identify the buggy model that predicts a wrong label. In the program~\cite{bug5}, the training data are in the range $[- \infty, + \infty ]$. This model has an issue that all the negative values are predicted to zeros, because the developer used the ReLU activation function in the last layer instead of activation functions that produce output in the same range of output labels such as TanH. 

In the program \cite{bug21},  \texttt{fit\_generator()} instead of \texttt{fit()} is used, which our technique has not yet supported. 
\texttt{fit\_generator()} is used to train the model on data generated batch-by-batch \cite{kerasDoc}. This API is not yet supported by our technique, we plan to cover and investigate other APIs in our future work.

In the \sof post \cite{bug24}, the user asked about the differences between two models, each one has different input dimensions, both models are training correctly, but the differences are in the performance. In this case, our tool will not find any numerical error or misbehavior to detect the bug.


In our tool, we did not target problems such as: (1) lack of dataset, (2) training dataset with distribution problems and (3) incorrect number of epochs, batch sizes, the number of hidden layers, and neural nodes in the layers. An example of these problems can be found in programs \cite{bug11, bug30}, these problems make the model terminates training at an early stage, and our tool will not detect the issue in the model.

 We also compared our approach against three \keras methods 
in terms of fault localization, We found that none of the three \keras approaches are able to determine the root causes. 
Our tool is able to determine root causes for 19 out 29 programs, 
and our callback is able to determine root causes for 21 out of 40 programs. 
We compared the results from our tool and our callback method with the 
ground truth we built for the benchmark. 
Our tool can precisely determine the layer or the parameter that causes 
the error for 19 out 29 cases and our callback method can precisely 
determine for 21 out 40 cases. 
In the rest of the models, the tool and the callback method reported 
the root cause in another layer because the training process is a cycle; 
the operation in one layer can affect the adjacent layer. 
For example, the program~\cite{bug3}, \texttt{mean\_absolute\_error} 
is used as a loss function instead of \texttt{mean\_squared\_error}, and 
the \texttt{SGD} optimizer is used instead of \texttt{Adam} optimizer. 
Our tool reports that the bug is EBW (error in weight in backward 
propagation) at layer. No = 1. 
In the program \cite{bug13}, the user assigned the learning rate = 0.1 
instead of learning rate = 0.001. 
Our tool reported EBA (Error before activation function), layer. No = 2.
\begin{figure}[!htb]
	\centering
	\includegraphics[width=3.5in,trim={1.0cm 3cm 0cm 1cm},clip]{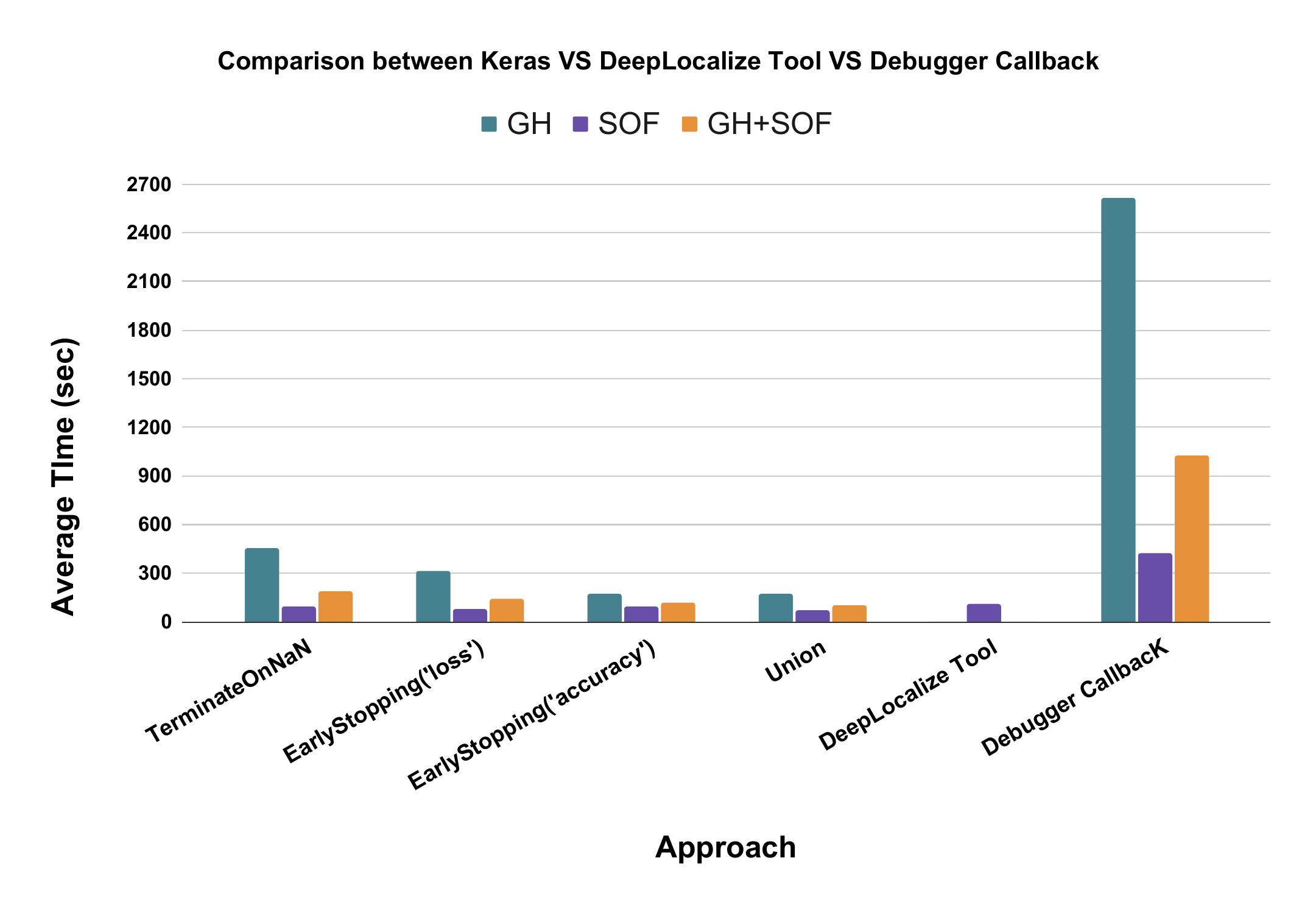}
	\caption{Comparison between \keras and Our Technique }
	\label{Comparsion}
\end{figure}

We measured the analysis time for the three \keras methods, our tool, 
and our callback function using \sof (SOF) and \gh (GH) benchmarks. 
Using SOF benchmarks,  \texttt{TerminateOnNaN(), EarlyStopping('loss'), EarlyStopping('accuracy')}, 
the three \keras methods together, our callback method and our tool take an average of
93.01, 78.98, 93.04, 73.55, 421.39, and 107.99 seconds respectively overall all \sof (SOF) benchmarks. 
While in GH benchmarks, \texttt{TerminateOnNaN(), EarlyStopping('loss'), EarlyStopping('accuracy')}, 
the three \keras methods together,  and our callback take an average of
451.52,	310.57,	171.22,	172.02, and 2613.6 seconds respectively.
The average time for all \sof (SOF) and \gh (GH) benchmarks is
191.6,	142.67,	114.54,	100.63, and 1024.25 seconds respectively for the four corresponding tools.  In \fignref{Comparsion}, we plot the average time, and shown that, \texttt{TerminateOnNaN()} terminates the program when there is a NaN in the loss value; in other cases, the training will continue. This is why it takes longer than other methods. Our callback function needs more value than the three \keras methods, such as update weight and gradient to perform dynamic analysis. This leads to an overhead if the model has a large number of parameters, but the overhead is usually insignificant compared to the amount of time that takes other callback functions. 
However, we anticipate that the time needed to manually debug and fault localization 
would be much more than the time required to execute our callback function. 

During building the benchmark from \sof and \gh and identifying the bugs using our technique, we observed that the most frequent bugs that the developer made are (1) choosing the activation, and loss function, as there are many combinations to choose the activation and loss depend on the type of the problem, (2) selecting the hyperparameters, such as the learning rate, the number of neural nodes in the layer, number of layers in the model and (3) preprocessing the training dataset, for example, the MNIST model needs to normalize the training dataset from the range [-256, + 256] to [-1, 1]. Without fault localization information, the developers will face the problem of determining the layers or parameters that induce the bug in the model.


\subsection{Summary}


\subsubsection{RQ1 (Validation)} 
We provide the empirical evidence (\tabref{tab:ImperativeVSKeras}) that 
training a DNN using our imperative program is consistent with using the \keras library. 
For RQ1, we show that our tool is able to identify 23 out of 29 faulty models, and our callback function is able to identify 34 out of 40 faulty models. 
Also our technique is able to determine the  fault localization precisely 
for 19 out 29 using our tool, and 21 out 40 using our callback 
while \keras is not able to localize bug. 

\subsubsection{RQ2 (Comparison)}
Our technique is able to identify 34 out of 40 faulty models using our 
callback functions, and 23 out of 29 using our tool. 
On the other hand, the \keras library reported bugs for 2, 24, and 28 out 
of 40 models, respectively. 
Also, if we take the union of three methods, the \keras library 
reported bugs for 32 out of 40 models. 
We also show that our technique can effectively locate the root 
causes for 21 out of 40 programs using our callback and 19 out 29 
using our tool while \keras methods does not support this feature. 
As it can be seen in \fignref{Comparsion}, our tool using dynamic 
analysis is practical and faster than \keras by identifying the bug. 
Three methods using \keras take on average 191.6, 142.67, and 114.54 
seconds respectively, while our tool takes on average 107.99 seconds. 
On the other hand, our callback needs more time to make dynamic 
analysis for additional parameters than loss and accuracy. 
This overhead is usually insignificant compared to the amount 
of time taken by \texttt{TerminateOnNaN()}.

\subsubsection{RQ3 (Limitation)} 
We failed to detect bugs for 7 out of 30 programs and failed to 
localize faults for 11 out of 30 programs using our tool, and 
our callback failed to localize fault for 20 out of 41. 
These programs either predicted wrong labels or have problems in the training dataset, the settings of epoches, batch sizes, the number hidden layers, and the number of neural nodes. 
In such cases, the training stopped before the model finished learning. 
Thus our tool was not able to handle these cases. 
Also, our prototype does not support all the \keras APIs. 
As an example,  one of the buggy program in \sof~\cite{bug21} 
used \texttt{fit\_generator()} instead of \texttt{fit()} function.  
\texttt{fit\_generator()} is used to train the model on data 
generated batch-by-batch \cite{kerasDoc}. 
We plan to cover and investigate other APIs in our future work.


\begin{table*}[htbp]
	
\centering
\caption{Comparisons Between the \keras Methods and Our Tool, and Our Callback Using \sof Post }
\scalebox{0.6}{
	\begin{tabular}{|r|r|r|r|r|r|r|r|r|r|r|r|r|r|r|r|r|r|r|r|r|r|r|r|r|r|}
		\hline
		\multicolumn{1}{|c|}{\multirow{2}[4]{*}{\textbf{Post \# $\mid$ Ref}}} & \multicolumn{5}{c|}{\textbf{TerminateOnNaN}} & \multicolumn{5}{c|}{\textbf{EarlyStopping(monitor='loss')}} & \multicolumn{5}{c|}{\textbf{EarlyStopping(monitor='accuracy')}} & \multicolumn{5}{c|}{\textbf{Our Tool}}   & \multicolumn{5}{c|}{\textbf{Our CallBack}} \\
		\cline{2-26}          & \multicolumn{1}{c|}{\textbf{Time}} & \multicolumn{1}{c|}{\textbf{Epoch}} & \multicolumn{1}{c|}{\textbf{Iteration}} & \multicolumn{1}{c|}{\textbf{IB}} & \multicolumn{1}{c|}{\textbf{FL}} & \multicolumn{1}{c|}{\textbf{Time}} & \multicolumn{1}{c|}{\textbf{Epoch}} & \multicolumn{1}{c|}{\textbf{Iteration}} & \multicolumn{1}{c|}{\textbf{IB}} & \multicolumn{1}{c|}{\textbf{FL}} & \multicolumn{1}{c|}{\textbf{Time}} & \multicolumn{1}{c|}{\textbf{Epoch}} & \multicolumn{1}{c|}{\textbf{Iteration}} & \multicolumn{1}{c|}{\textbf{IB}} & \multicolumn{1}{c|}{\textbf{FL}} & \multicolumn{1}{c|}{\textbf{Time}} & \multicolumn{1}{c|}{\textbf{Epoch}} & \multicolumn{1}{c|}{\textbf{Iteration}} & \multicolumn{1}{c|}{\textbf{IB}} & \multicolumn{1}{c|}{\textbf{FL}} & \multicolumn{1}{c|}{\textbf{Time}} & \multicolumn{1}{c|}{\textbf{Epoch}} & \multicolumn{1}{c|}{\textbf{Iteration}} & \multicolumn{1}{c|}{\textbf{IB}} & \multicolumn{1}{c|}{\textbf{FL}} \\
			\hline
		\rowcolor[rgb]{ .859,  .859,  .859} 48385830 $\mid$ \cite{bug1} & 1.20  & 1     & 10    & \greencheck   & --  & 12.21 & 1     & 60000 & \greencheck   & --  & 34.90 & 3     & 60000 & \greencheck   & --  & 0.14  & 1     & 4     & \greencheck   & \greencheck & 2.14&	1&	20 &\greencheck   & \greencheck \\
		\hline
		44164749 $\mid$ \cite{bug2} & 175.23 & 5     & 60000 & X   & -- & 175.23 & 5     & 60000 & X   & --  & 175.23 & 5     & 60000 & X   & --  & 22.93&	1&	50 & \greencheck   & \greencheck & 111.56&	1	&2000&  \greencheck   & \greencheck \\
		\hline
		\rowcolor[rgb]{ .859,  .859,  .859}31556268 $\mid$ \cite{bug3} & 5.69  & 1000  & 4     & X   & -- & 5.69  & 1000  & 4     & X   & -- & 0.91  & 2     & 4     & \greencheck   & -- & 4.57  & 98    & 394   & \greencheck   & X & 1.20&	1&	4& \greencheck & X\\
		\hline 
		50306988 $\mid$ \cite{bug4} & 1.44  & 5     & 200   & X   & --  & 1.44  & 5     & 200   & X   & --  & 1.30  & 2     & 200   & \greencheck   & --  & 1.03  & 1     & 100   & \greencheck   & \greencheck & 3.57&	1&	64 &\greencheck   & \greencheck \\
		\hline
		\rowcolor[rgb]{ .859,  .859,  .859}48251943 $\mid$ \cite{bug5} & 5.22  & 100   & 17    & X   &  -- & 0.93  & 27    & 17    & \greencheck   & -- & 5.22  & 100   & 17    & X   &  -- & 20.45 &	100 &	17    & X   & X & 706.83&	100&	17 & X & X\\
		\hline
		38648195 $\mid$ \cite{bug6} & 120.52 & 20    & 48000 & X   & --  & 120.52 & 20    & 48000 & X   & --  & 120.00 & 20     & 48000 & X   & --  & 2.49  & 1     & 95    & \greencheck   & X & 25.92&	1&	100 & \greencheck   & \greencheck\\
		\hline
		\rowcolor[rgb]{ .859,  .859,  .859}33969059 $\mid$ \cite{bug7} & 130.66 & 20    & 10000 & X   & --  & 14.03 & 2     & 10000 & \greencheck   & --  & 14.14 & 2     & 10000 & \greencheck   & --  & 0.95  & 1     & 50    & \greencheck   & \greencheck & 1.52&	1&	32 & \greencheck   & \greencheck\\
		\hline
		55328966 $\mid$ \cite{bug8} & 98.84 & 10    & 49999 & X   & --  & 98.84 & 10    & 49999 & X   & --  & 98.84 & 10    & 49999 & X   & --  & 272.70 & 1     & 737   & \greencheck   & \greencheck &9092.93&	10&	49999 & X   & X\\
		\hline
		\rowcolor[rgb]{ .859,  .859,  .859}34311586 $\mid$ \cite{bug9} & 1.01  & 20    & 6     & X   & --  & 0.87  & 2     & 6     & \greencheck   & --  & 0.97  & 20    & 6     & X   & --  & 0.73  & 8     & 50    & \greencheck   & \greencheck & 1.02&	1&	32& \greencheck   & \greencheck\\
		\hline
		31880720 $\mid$ \cite{bug10} & 176.02 & 3     & 20000 & X   & --  & 110.25 & 2     & 20000 & \greencheck   & --  & 113.28 & 2     & 20000 & \greencheck   & --  & 1.11  & 1     & 50    & \greencheck   & \greencheck & 2.12	& 1 &	1 & \greencheck   & \greencheck\\
		\hline
		\rowcolor[rgb]{ .859,  .859,  .859}39525358 $\mid$ \cite{bug11} & 3.15  & 150   & 13    & X   & -- & 3.15  & 150   & 13    & X   & -- & 1.68  & 2     & 13    & \greencheck   & -- & 28.36 &	150	& 13 & X   & X & 14.29	&2&	10  &\greencheck   & \greencheck\\
		\hline
		39217567 $\mid$ \cite{bug12} & 16.97 & 1000  & 256   & X   & --  & 1.31  & 11    & 256   & X   &-- & 1.18  & 2     & 256   & \greencheck   & --  & 329.46	&100	&256   & X   & X & 188.71 &	5 &	50 & \greencheck   & X\\
		\hline
		\rowcolor[rgb]{ .859,  .859,  .859}48934338 $\mid$ \cite{bug13} & 11.03 & 1000  & 50    & X   & -- & 0.91  & 10    & 50    & \greencheck   & -- & 0.85  & 2     & 50    & \greencheck   & -- & 0.33  & 1     & 13    & \greencheck   & X & 3.01&	1&	32 &  \greencheck   & X\\
		\hline
		47724077 $\mid$ \cite{bug14} & 31.07 & 50    & 32561 & X   & --  & 2.92  & 3     & 32561 & \greencheck   & --  & 3.46  & 4     & 32561 & \greencheck   & --  & 0.66  & 1     & 50    & \greencheck   & \greencheck & 1.42	&1&	100 & \greencheck   & \greencheck  \\
		\hline
		\rowcolor[rgb]{ .859,  .859,  .859} 59325381 $\mid$ \cite{bug15} & 758.04 & 10    & 60000 &     X  &  --     & 260.06 & 3     & 60000 &    \greencheck   &     --  & 239.52 & 3     & 60000 &    \greencheck   &   --    &      125.52 &     	1	  &   1750    &     \greencheck  & X & 24.66	& 1&	160 & \greencheck   & \greencheck \\
		\hline
		59278771 $\mid$ \cite{bug16} & 1.94  & 200   & 135   & X   & --& 5.24  & 62    & 135   & \greencheck   & -- & 1.94  & 200   & 135   & X   & -- & 0.52	& 1 &	50   & \greencheck   & \greencheck & 60.64	& 1 &	75& \greencheck   & X \\
		\hline
		\rowcolor[rgb]{ .859,  .859,  .859} 52800582 $\mid$ \cite{bug17} & 10.04 & 1000  & 400   & X   & --  & 1.79  & 2     & 400   & \greencheck   & --  & 10.04 & 1000  & 400   & X   & --  & 0.14  & 1     & 7     & \greencheck   & \greencheck & 12.79	& 4&	400 & \greencheck   & X\\
		\hline
		41372874 $\mid$ \cite{bug18} & 6.86  & 20    & 239   & X   & -- & 6.86  & 20    & 239   & X   & -- & 2.40  & 3     & 239   & \greencheck   & -- & 66.65 &	20 &	239   & X   & X & 1.42 &	1 &	2 & \greencheck   & \greencheck\\
		\hline
		\rowcolor[rgb]{ .859,  .859,  .859}34673164 $\mid$ \cite{bug19} & 1.93  & 9     & 20    & X   & --  & 2.00  & 2     & 20    & \greencheck   & --  & 2.09  & 6     & 20    & \greencheck   & --  & 2.04 &	11	& 100    & \greencheck   & \greencheck & 26.00	& 5 &	9 & \greencheck   & \greencheck\\
		\hline
		48221692 $\mid$ \cite{bug20} & 7.16  & 1000  & 150   & X   & --  & 2.61  & 280   & 150   & \greencheck   & --  & 0.95  & 2     & 150   & \greencheck   & --  & 24.24 & 13    & 150  & \greencheck   & \greencheck & 8.02 &	4 &	150 & \greencheck & X \\
		\hline
		\rowcolor[rgb]{ .859,  .859,  .859} 50079585 $\mid$ \cite{bug21} & 171.69 & 10    & 61    &  X     &  --     & 172.00 & 10    & 61    & X      &  --     & 54.38 & 3     & 61    &   \greencheck    &  --     & --       &    --    &     --   &  --      & --  & -- & -- & -- & -- &\\
		\hline
		45337371 $\mid$ \cite{bug22} &      12.52	& 5	      &    1000	   &  X     &  --     &   6.79	    &    2  &     	1000	   & \greencheck      &    --   & 7.02	      &   2	    &   1000    &   \greencheck    &  --     &    1.17 &	1	& 50  & \greencheck   & \greencheck & 2.43	&1&	1 & \greencheck      &\greencheck  \\
		\hline
		\rowcolor[rgb]{ .859,  .859,  .859}44066044 $\mid$ \cite{bug23} & 3.87  & 100   & 21    &   X    &  --     & 2.07  & 4     & 21    &  \greencheck     &    --   & 1.97  & 2     & 21    &  \greencheck     &     --  &    1.07	& 2	& 50 & \greencheck      &\greencheck  & 1.41 &	1&	5& \greencheck      &\greencheck  \\
		\hline
		51930566 $\mid$ \cite{bug24} & 4.07  & 50    & 75    & X   & -- & 4.07  & 50    & 75    & X   & -- & 1.66  & 2     & 75    & \greencheck   & -- & 38.65  & 50    & 75    & X   & X & 22.40&	1&	40 & \greencheck   & X \\
		\hline
		\rowcolor[rgb]{ .859,  .859,  .859} 47352366 $\mid$ \cite{bug25} & 32.75      &  1     &  4950  &   \greencheck    &  --     &    381.31   &    1   & 60000
		&   \greencheck    &  --     &     793.64	& 2	 &  60000 &   \greencheck    &  --      &  75.46 &	1 &	50 &\greencheck
		& \greencheck & 1177.03&	1&	1050 & \greencheck      &  X \\
		\hline
		45442843 $\mid$ \cite{bug26} & 2.57  & 100   & 200   & X   & --  & 1.17  & 3     & 200   & \greencheck   & --  & 1.15  & 2     & 200   & \greencheck   & --  & 0.48  & 1     & 50    & \greencheck   & \greencheck & 0.97	&1&	200 & \greencheck      &  \greencheck \\
		\hline
		\rowcolor[rgb]{ .859,  .859,  .859}48594888 $\mid$ \cite{bug27} & 1055.29 & 5     & 50000 &   X    &  --     & 1055.29 & 5     & 50000 &    X   &  --     & 1055.29 & 5     & 50000 &    X   &  --     &   116.38    &   1   &  51      & \greencheck      &  \greencheck & 4.24 & 1 & 32 & \greencheck      &  \greencheck \\
		\hline
		31627380 $\mid$ \cite{bug28} & 6.28  & 10    & 712   & X   & -- & 6.12  & 10    & 712   & X   & --  & 3.09  & 3     & 712   & \greencheck   & --  & 1.28  & 1     & 50    & \greencheck   & \greencheck & 1.76	&1&	4& \greencheck      &  \greencheck \\

		\hline
		\rowcolor[rgb]{ .859,  .859,  .859}58609115 $\mid$ \cite{bug29} & 3.12  & 10    & 1767  & X   & --  & 2.61  & 8     & 1767  & \greencheck   & --  & 1.94  & 3     & 1767  & \greencheck   & --  & 29.06 & 1     & 1350  & \greencheck   & \greencheck & 343.78 &	1&	1250& \greencheck      &  X \\
		\hline
		50481178 $\mid$ \cite{bug30} & 12.66 & 50    & 200   & X   & -- & 4.12  & 6     & 200   & \greencheck   & -- & 3.58  & 3     & 200   & \greencheck   & -- & 1963.13&	50&	200   & X   & X & 376.44 &	1 &	150 & \greencheck      &  X \\
		\hline
		
	\end{tabular}}%
\label{tab:Comparison}%
\end{table*}%

\begin{table*}[htbp]
	
\centering
\caption{Comparisons Between the \keras Methods and Our Callback Using \gh Repository }
\scalebox{0.61}{
	\begin{tabular}{|r|r|r|r|r|r|r|r|r|r|r|r|r|r|r|r|r|r|r|r|r|r|r|r|r|r|}
		\hline
		\multicolumn{1}{|c|}{\multirow{2}[4]{*}{\textbf{GH $\mid$ Ref}}} & \multicolumn{5}{c|}{\textbf{TerminateOnNaN}} & \multicolumn{5}{c|}{\textbf{EarlyStopping(monitor='loss')}} & \multicolumn{5}{c|}{\textbf{EarlyStopping(monitor='accuracy')}} & \multicolumn{5}{c|}{\textbf{Union}}   & \multicolumn{5}{c|}{\textbf{Our CallBack}} \\
		\cline{2-26}          & \multicolumn{1}{c|}{\textbf{Time}} & \multicolumn{1}{c|}{\textbf{Epoch}} & \multicolumn{1}{c|}{\textbf{Iteration}} & \multicolumn{1}{c|}{\textbf{IB}} & \multicolumn{1}{c|}{\textbf{FL}} & \multicolumn{1}{c|}{\textbf{Time}} & \multicolumn{1}{c|}{\textbf{Epoch}} & \multicolumn{1}{c|}{\textbf{Iteration}} & \multicolumn{1}{c|}{\textbf{IB}} & \multicolumn{1}{c|}{\textbf{FL}} & \multicolumn{1}{c|}{\textbf{Time}} & \multicolumn{1}{c|}{\textbf{Epoch}} & \multicolumn{1}{c|}{\textbf{Iteration}} & \multicolumn{1}{c|}{\textbf{IB}} & \multicolumn{1}{c|}{\textbf{FL}} & \multicolumn{1}{c|}{\textbf{Time}} & \multicolumn{1}{c|}{\textbf{Epoch}} & \multicolumn{1}{c|}{\textbf{Iteration}} & \multicolumn{1}{c|}{\textbf{IB}} & \multicolumn{1}{c|}{\textbf{FL}} & \multicolumn{1}{c|}{\textbf{Time}} & \multicolumn{1}{c|}{\textbf{Epoch}} & \multicolumn{1}{c|}{\textbf{Iteration}} & \multicolumn{1}{c|}{\textbf{IB}} & \multicolumn{1}{c|}{\textbf{FL}} \\
			\hline
	    1 $\mid$ \cite{GitHub2}  & 356.14 &	2	& 1944601	& X	& --	& 345.82	& 2	& 1944601	& X	& -- & 	351.46	& 2	& 1944601 &	X & 	--	& 343.17 & 	2	& 1944601 &	X &	-- &11.80&	1&	96 &	\greencheck &	X\\
		\hline
		\rowcolor[rgb]{ .859,  .859,  .859} 2 $\mid$ \cite{GitHub3}  & 3427.71 &	50	& 100	& X	& -- &	2602.44 &	36 &	100 &	\greencheck &	-- &	1082.78 &	15 &	100 &	\greencheck & 	-- &	1073.39 & 	15 &	100 &	\greencheck &	-- & 8432.06 &	50 &	100 &	X &	X\\
		\hline
		 3 $\mid$ \cite{GitHub5} & 101.78 &	100 &	100 &	X &	-- &	9.01 &	8 &	2140 &	\greencheck &	-- &	4.39 &	3	 & 2140	 & \greencheck & 	-- &	7.20 &	5	& 2140 &	\greencheck &	-- & 31.69&	1&	96&	\greencheck &	\greencheck \\
		\hline
		\rowcolor[rgb]{ .859,  .859,  .859} 4 $\mid$ \cite{GitHub6}  & 256.76 &	20 &	1999 &	X &	-- &	52.46 &	4	& 1999	 & \greencheck &	-- &	54.10 &	4	& 1999 &	\greencheck & 	-- &	52.88 &	20 &	1999 & \greencheck & -- & 102.44 &	1 &	300 & 	 \greencheck  & X\\
		\hline
		5 $\mid$ \cite{GitHub7}  & 168.86 &	10	& 27839 & 	X & --	 & 169.24 &	10 &	27839 &	X &	-- &	171.04 &	10 &	27839 &	X &	-- &	191.71 &	10 &	27839 &	X &	-- & 164.70	& 1 &	2560 &	\greencheck &	\greencheck \\
		\hline
		\rowcolor[rgb]{ .859,  .859,  .859} 6 $\mid$ \cite{GitHub11} & 19.92 &	25	& 768 &	X &	-- &	2.44 &	10 &	768 &	\greencheck &	--	& 1.54 &	2 &	768 &	\greencheck &	-- &	1.52 &	2	& 768 &	\greencheck & -- & 9568.09 &	25 &	768 &	X & 	X\\
		\hline
		7 $\mid$ \cite{GitHub17}  & 19.33	& 10 &	1166 &	X &	-- &	19.75 &	10	& 1166 &	X &	-- &	4.90	& 3	& 1166 &	\greencheck &	-- &	4.95 & 3 &	1166 & \greencheck &	-- & 1.90 &	2 &	1 &	\greencheck & 	\greencheck\\
		\hline
		\rowcolor[rgb]{ .859,  .859,  .859} 8 $\mid$ \cite{GitHub22} & 2.66 &	2	&125 &	X	& -- &	2.51 &	2	& 125 &	X & -- &	2.74 &	2	& 125 &	X &	-- &	2.58 &	2	& 125 & X &	-- & 1022.32&	2&	125	 & X & X\\
		\hline
		9 $\mid$ \cite{GitHub23}  & 13.30 &	4 &	15000 &	X	& -- &	12.67 &	4 &	15000 &	X &	-- &	13.55 &	4 &	15000 &	X &	-- &	13.38 &	4 &	15000 &	X &	-- & 9381.21&	4&	15000 &	X &	X\\
		\hline
		\rowcolor[rgb]{ .859,  .859,  .859} 10 $\mid$ \cite{GitHub25} & 587.26 &	15 &	1440 & X	& -- &	194.23 &	5	& 1440 &	\greencheck &	-- &	193.74 &	5	& 1440 & 	\greencheck &	-- & 	198.24 &	5 &	1440 &	\greencheck & -- & 28.04	& 1 &	64 &	\greencheck &	\greencheck \\
		\hline
		11 $\mid$ \cite{GitHub27}   & 12.97 &	10 &	10000 &	X &	--	& 5.73 &	4	& 10000 &	\greencheck &	--	 & 3.20 &	2	 & 10000 &	\greencheck &	-- &	3.23 &	2	& 10000 &	\greencheck &	-- & 5.38 &	1&	30 & 	\greencheck & 	X\\
		\hline
		
	\end{tabular}}%
\label{tab:ComparisonGitHub}%
\end{table*}%

%% file: threats.tex
\section[Threats]{Threats to Validity}
\label{sec:THREATSTOVALIDITY}

We were mainly concerned with the implementation of our tool that may affect the evaluation and result. To minimize this threat, we followed \keras code \cite{kerasCode} and carefully reviewed our implementation between the authors to reduce the chances that major errors were compared to our tool against \keras, as we have shown in Table \ref{tab:ImperativeVSKeras}.

Our tool currently handles the frequent layers/APIs in the \keras library, including:  \texttt{Dense}, \texttt{Dropout}, \texttt{MaxPooling2D}, \texttt{Conv2D}, and \texttt{BatchNormalization}. We may not yet able to handle deep learning programs whose bugs are related to other APIs e.g., \texttt{ConvLSTM2D()}, \texttt{Conv3D()}.

We validated the output of our tool using the benchmarks we created. The trustworthy of manually created benchmarks could be a threat to the validity of our results. To alleviate this threat, we follow the same methodology in prior works \cite{jones2002visualization}. Also, we have evaluated the benchmark before/after fixing the issue in the model.


%% file: related.tex
\section[Related]{Related Work}
\label{sec:relatedwork}

The closest related work in terms of technical ideas is by 
Gopinath~\etal~\cite{gopinath2019symbolic1, gopinath2019symbolic2}. 
Gopinath~\etal proposed a new approach (DeepCheck) inspired 
from program analysis to test a Deep Neural Network (DNN) using symbolic execution.  
DeepCheck also uses a white box technique to enable symbolic execution 
to find the important pixels, and find the attack pixels by translating 
DNN to an imperative program that has the same behavior as DNN. 
The experimental results conducted using MNIST data set shows that their 
approach is able to create 1-pixel and 2-pixel attacks by finding the most 
important pixels that DNN fails to classify correctly.
Our imperative representation is inspired by this work. 
While this work focuses on identifying adversarial attack, our work
aims to identify faults in DNNs and localize their fault. 

\subsection{Testing Deep Neural Networks}
DeepTest~\cite{DeepTest} and follow-up work aim to automatically generate
test cases to examine corner cases corresponding to real world inputs. 
These works have focused on test case generation, whereas our work uses
an existing training dataset and focuses on localizing the root cause of 
a bug by observing the runtime behavior of a deep neural network. 
Zhang~\etal~\cite{zhang2020machine} present a comprehensive 
survey of work in this area. 
Eniser~\etal~\cite{DeepFault} proposed a new white box analysis technique (DeepFault) inspired from spectrum-based fault localization.  DeepFault tests DNN models to achieve two objectives: 
(i) detect suspicious neurons, i.e., neurons likely to be more responsible for inadequate DNN performance; and (ii) synthesis of new inputs, using correctly classified inputs, that exercise the identified suspicious neurons. The experimental results conducted on MNIST and CIFAR-10 datasets show that DeepFault is effectively identifying suspicious neurons. DeepFault does not focus on structure bugs, instead they focus on training bug, and pre-trained DNNs analysis, given a specific test set. While our tool identifies the buggy model and faults the root causes of structure bug in DNN.

\subsection{Empirical Study on Deep Learning Bugs} 
There have been several empirical studies that have analyzed different 
kind of bugs in deep learning networks. 
These studies have been conducted on real code and examples from the 
\sof posts and \gh issues. They have focused on symptoms and 
root causes of bugs to have a better understanding of deep learning bugs. 

Zhang \etal~\cite{zhang2019empirical} utilized \sof posts 
and \gh commits to investigate bugs in deep learning applications 
built on top of TensorFlow. They focused on symptoms and root causes of TensorFlow bugs to have a better understanding of deep learning bugs.  
Islam \etal~\cite{islam2019comprehensive} also studied deep learning bugs using 
\sof questions and \gh commits. 
They also adapted a taxonomy of bug type, root cause, and bug impacts 
for deep learning software for five popular deep learning libraries. 

Another study has been conducted to understand the bug fix pattern and how 
the developer can develop a tool to fix the bugs automatically by Islam \etal~\cite{islam20repairing}. They have conducted a comprehensive study on 415 posts from \sof, and 555 from \gh commits for five popular DNN packages to understand the bug fix patterns, and how the bugs can be fixed in DNN software. The main goal of this study to help the developers to understand the characteristics of bug and how they can design an automated repair tool.

\subsection{Bugs Repairing in Deep Learning} 
In recent years, there are several proposals for debugging deep neural networks. These techniques are often inspired from software debugging and testing techniques. 

Ma \etal~\cite{ma2018mode} proposed and developed a technique called MODE inspired by software debugging. MODE performs state differential analysis to solve two types of problems: overfitting problems and under-fitting problems. MODE can solve these problems by identify the buggy features (or neurons) that are responsible for the misclassification in the model, then it constructs the degree of importance of features to retrain the faulty neurons with new input samples selection. MODE provides effective and efficient method to fix the buggy models without introducing new bugs, and compromise on accuracy and training time cost.

Zhang \etal~\cite{zhang2019apricot} introduced an automatic approach to fixing deep learning models called Apricot. Apricot is able to adjust the ill-trained weights without using additional training data or any artificial parameters, Apricot using a set of reduced models from the original model, and compare the differences between the original model and correct/incorrect of reduced models iteratively, to find these failing test case that are responsible for the misclassification in the original model. The approach uses three strategies to adjust the weight and achieve a higher test accuracy. The experimental results using CIFAR-10 dataset and five state-of-the-art of deep learning models have shown that the approach can increase the test and training accuracy.

In recent years, several researchers are supporting automated debugging and repair approaches for deep neural networks, and recent research is summarized in \cite{zhang2020machine}. This topic is still at the early stages~\cite{ma2018mode, zhang2019apricot}.  To the best of our knowledge, all previous works are focused on the training bugs. 
Our technique is the first approach that automatically identifies 
the buggy model and localizes the root causes of structure bug in 
DNNs by applying the DNN Bug Detection algorithm. 

%% file: conclusion.tex
\section[Conclusion]{Conclusions and Future Work}
\label{sec:conclusion}
As deep neural networks are becoming integrated with software systems 
from different domains, 
developing debugging techniques to localize the root cause of the bug has
become urgent. 
Thus, recent work has developed techniques to inspect the entire model and find faults. 
Inspired by program analysis and software debugging techniques, we have presented 
an automated approach powered by a dynamic analysis and statistical analysis. 
It can help identify the buggy model and the root causes of DNN error. 
An experimental evaluation using 40 real deep learning applications 
shows the usefulness of our technique.
For 34/40 cases, our approach was able to detect faults whereas the 
best debugging approach provided by \keras detected 32/40 faults.
Our approach was able to localize 21/40 bugs whereas 
\keras did not localize any faults.

Future work includes developing techniques to repair deep 
neural network bugs, and exploring cases that our work was 
unable to detect faults (6/40) and localize errors (19/40). 
Recent work has also used analysis of the DNN structure to 
decompose it into modules~\cite{pan20decomposing}. 
It would be interesting to explore whether a similar mechanism can be
utilized for better localization. 
It would also be interesting to go beyond accuracy bugs to detect and localize more 
non-functional bugs, e.g. fairness bugs~\cite{biswas20machine}.


%% file: acknowledgment.tex
\section[Acknowledgment]{Acknowledgment}
\label{sec:Acknowledgment}
This work was supported in part by the US National Science Foundation (NSF) under grants CNS-15-13263, and CCF-19-34884. 
All opinions are of the authors and do not reflect the view of sponsors.
This work benefited from discussions with Hamid Bagheri and Johirul Islam.